\documentclass[twocolumn,prb]{revtex4-1}
\usepackage{amsmath, amssymb, braket, dsfont, units,color}
\usepackage{graphicx}
\usepackage{subfigure}		
\usepackage{epstopdf, times}
\usepackage[utf8]{inputenc}

\begin{document}

\title{Absence of Many-Body Localization in a Single Landau Level}

\author{Scott D. Geraedts and R. N. Bhatt}
\affiliation{Department of Electrical Engineering, Princeton University, Princeton NJ 08544, USA}
\begin{abstract}
The localization properties of the highly excited states of a system projected into a single Landau level are analyzed. 
An analysis of level spacing ratios for finite size systems shows a clear crossover from extend (GUE) to localized (Poisson) statistics, indicating a many body localization transition.
However, the location of this transition depends very strongly on system size, and appears to scale to infinite disorder  in the thermodynamic limit. 
This result does not depend on the properties of the ground state (such as whether the ground state exhibits topological order), as expected for a transition of highly-excited eigenstates.
We therefore conclude that many body localization does not exist in these systems.
Our results demonstrate that a sub-thermodynamic number of single particle effectively extended states is sufficient to cause all many body states to become extended. 
\end{abstract}
\maketitle

\section{Introduction}


An open question in the study of many-body localization is what adding interactions does to a system which has both localized and extended states in the non-interacting limit. 
Initial work\cite{Rahul_bath2014,Sarang,Johri_bath} showed that a thermal bath will thermalize a many-body localized system, even when the coupling between the bath and the system is small. Later work\cite{Rahul_bath2015} showed that when the system is allowed to affect the bath, it is possible for the system to localize the bath. 

A related problem is whether it is possible for a system to have a many-body mobility edge, i.e. to have both excited and localized states in its many body energy spectrum. Early numerical work \cite{Alet2015} showed the existence of such a phenomenon at intermediate disorder strengths, essentially due to the fact that the localization transition happens at different disorder strengths as a function of energy density. Other works\cite{Li2016,Modak2016} have presented numerical evidence that if one starts with a system with a single-particle mobility edge (i.e. the single particle spectrum has a finite fraction of extended states), adding interactions can give a many-body mobility edge. On the other hand, there are arguments that such many-body mobility edges are impossible\cite{bubbles}.

A way to shine light on these issues is to look at systems in a single Landau level. One motivation is that the new quantum technologies which first motivated the study of many-body localization are now being applied to Landau level physics \cite{Simon,Hafezi}. Additionally, in a Landau level the localization length of single-particle states necessarily diverges at a critical energy \cite{Halperin, Huckestein-Kramer, Huo-Bhatt}; this implies that at any finite length scale $L$, there are single particle states that are effectively extended. If we think of these states as a delocalized bath, we can ask how a localized system responds to a bath which makes up only a vanishing number of states in the thermodynamic limit \cite{Huo-Bhatt}, but which cannot be localized due to the topological nature of the Landau level. One might think that this makes a single Landau level a good candidate when searching for a many-body mobility edge. The single particle spectrum has very few extended states, and therefore in the absence of topological protection one might expect that the entire spectrum would become localized. The topological character of a single Landau level prevents this, but one may suspect that there are still not enough single particle extended states to delocalize the entire system, and therefore there may be a many-body mobility edge.

This question was approached analytically by Nandkishore and Potter,\cite{PotterNandkishore} who argued that the presence of single-particle extended states can delocalize the entire many-body spectrum. They studied systems with `marginal localization', like the Landau level with disorder, where the single particle spectrum has a divergent correlation length at a specific energy, and the correlation length diverges with critical exponent $\nu$. The fraction of `effectively extended states' at a length scale $L$ in the single particle spectrum is $\propto L^{ - 1/\nu}$ \cite{Huo-Bhatt, Yang-Bhatt}. They find that the spectrum is entirely delocalized whenever $\nu d >1$, a criterion that is certainly satisfied in the quantum Hall (single Landau level) problem for non-interacting electrons in a disorder potential where $\nu$ has been found to be around $2.5$ \cite {Huckestein-Kramer, Huo-Bhatt, Slevin-Ohtsuki, Obuse-Evers, Wan-Bhatt}. Further, $\nu d \ge 2$ is argued to hold more generally for transitions in disordered systems\cite{CCFS}. This finding poses a challenge to the findings of Refs.~\onlinecite{Li2016, Modak2016}, since if even a \emph{vanishingly small} fraction of extended states is sufficient to delocalize the entire system it might seem unlikely that many body mobility edges can exist when the fraction of extended states is nonzero in the thermodynamic limit. Nonetheless, there are some loopholes in the arguments of Ref.~\onlinecite{PotterNandkishore}. In particular, what the authors show is not that the system delocalizes when $\nu d >1$, but simply that perturbation theory breaks down. This motivates a numerical test of the findings of Ref.~\onlinecite{PotterNandkishore}, which is the subject of this work.

In this work we perform exact diagonalization on a single Landau level with interactions and Gaussian disorder. We measure the level spacing ratio of the resulting eigenvalue spectrum to determine whether the system experiences many-body localization, finding that for finite-size systems the system is localized at large disorder. However, the critical disorder at which the system localizes increases rapidly as a function of system size, leading us to conclude that in the thermodynamic limit there is no localization. Specifically, we find that the critical disorder strength increases as a power law in system size. This is consistent with the findings of Ref.~\onlinecite{PotterNandkishore}, though we observe a much larger power law exponent than they predict.

\section{Model and Method}
The Hamiltonian we study numerically is as follows:
\begin{equation}
H=P_{LLL} \left[ \sum_{i<j}^{N} V(\vec r_i-\vec r_j) + \sum_i^N U(\vec r_i)\right],
\label{hamiltonian}
\end{equation}
where $\vec r_i$ are the positions of the $N_e$ electrons in the problem. $V(\vec r)$ is the electron-electron interaction; for this work we will use a delta function interaction (which corresponds to Haldane pseudopotential $V_1$). The strength of the interaction is proportional to $e^2/\ell_B$, where $\ell_B$ is the magnetic length. 
$U(\vec r_i)$ is an on-site disorder potential with the property\cite{HuckesteinRMP}
\begin{equation}
\langle U(\vec r) \rangle=0, ~~~ \langle U(\vec r)U(\vec {r^\prime}) \rangle=W^2\delta(\vec r - \vec{r^\prime}). 
\end{equation}
The disorder potential is specified in momentum space, with a cutoff at large wavevectors. When the disorder potential is written in terms of the matrix elements of single-particle orbitals, terms at large momenta are exponentially suppressed. Therefore we choose a cutoff such that these matrix elements are below machine precision.
We will be studying systems where the disorder strength, $W$, is much larger than the interaction strength. This is unlike previous studies of this Hamiltonian\cite{Wan2005,Sheng2005,Liu2016} where the focus was on the ground state transition out of the fractional quantum Hall phase. Here we are interested in the eigenstate phase transition which occurs for the excited states of the system, and this transition (if it exists) would be expected to take place at much larger disorder strengths.

The operator $P_{LLL}$ projects this Hamiltonian into the lowest Landau level. Therefore in this problem we are studying the regime $h\omega_c \gg W \gg e^2/\ell_B$, where $h\omega_c$ is the spacing between Landau levels. This regime would only be accessible experimentally in the limit of very large magnetic fields.

We diagonalize the above Hamitonian for systems with $9-20$ flux quanta, accessing excited states using the shift-invert method\cite {referenceSI}. The Hilbert space sizes we can access are small compared to the sizes that be accessed for spin chains because the Hamiltonian in Eq.~(\ref{hamiltonian}) is considerably less sparse than a spin chain Hamiltonian. In the shift-invert method, one computes a decomposed version of the Hamiltonian matrix and uses this decomposed version to repeatedly solve a linear system \cite{referenceSI}. For the largest system sizes we study, our decomposed matrix is $\approx 50$ times less sparse than a spin chain Hamiltonian with similar Hilbert space dimension.

We diagnose localization by looking at the spacings of the energy levels, which are expected to follow a generalized unitary ensemble (GUE) distribution in the delocalized phase and a Poisson distribution in the localized case. The distribution for any size and disorder can be concisely characterized using the parameter $r$, defined as:
\begin{equation}
r\equiv \sum_i \frac {min (E_i-E_{i-1},E_{i+1}-E_i) }{ max (E_i-E_{i-1},E_{i+1}-E_i)},
\end{equation}
$r$ takes the value $0.60$ for a GUE distribution and $0.396$ for a Poisson distribution \cite{PalHuse}. 

For the non-interacting case, the many-body eigenstates are Slater determinants of $N_e$ single particle eigenstates that are, at high energy density, selected essentially at random. The total energy of a many-body state is the sum of the energies of the single particle states, the vast majority of which are localized. Two many-body states which are adjacent in energy obviously contain different occupied single-particle states. If two states have all the same localized single-particle states, and therefore differ only in the occupation of extended states, then those many-body states will exhibit level repulsion. However, if the many body states have different occupations of localized states, then there will be no level repulsion. Since most single-particle states are localized we expect that in the non-interacting limit, there will be no level repulsion and the level spacings should follow a Poisson distribution. As interactions are added, the distribution will stay Poisson unless the interactions are strong enough to delocalize all of the states.

\section{Results}

\begin{figure}
\includegraphics[width=\linewidth]{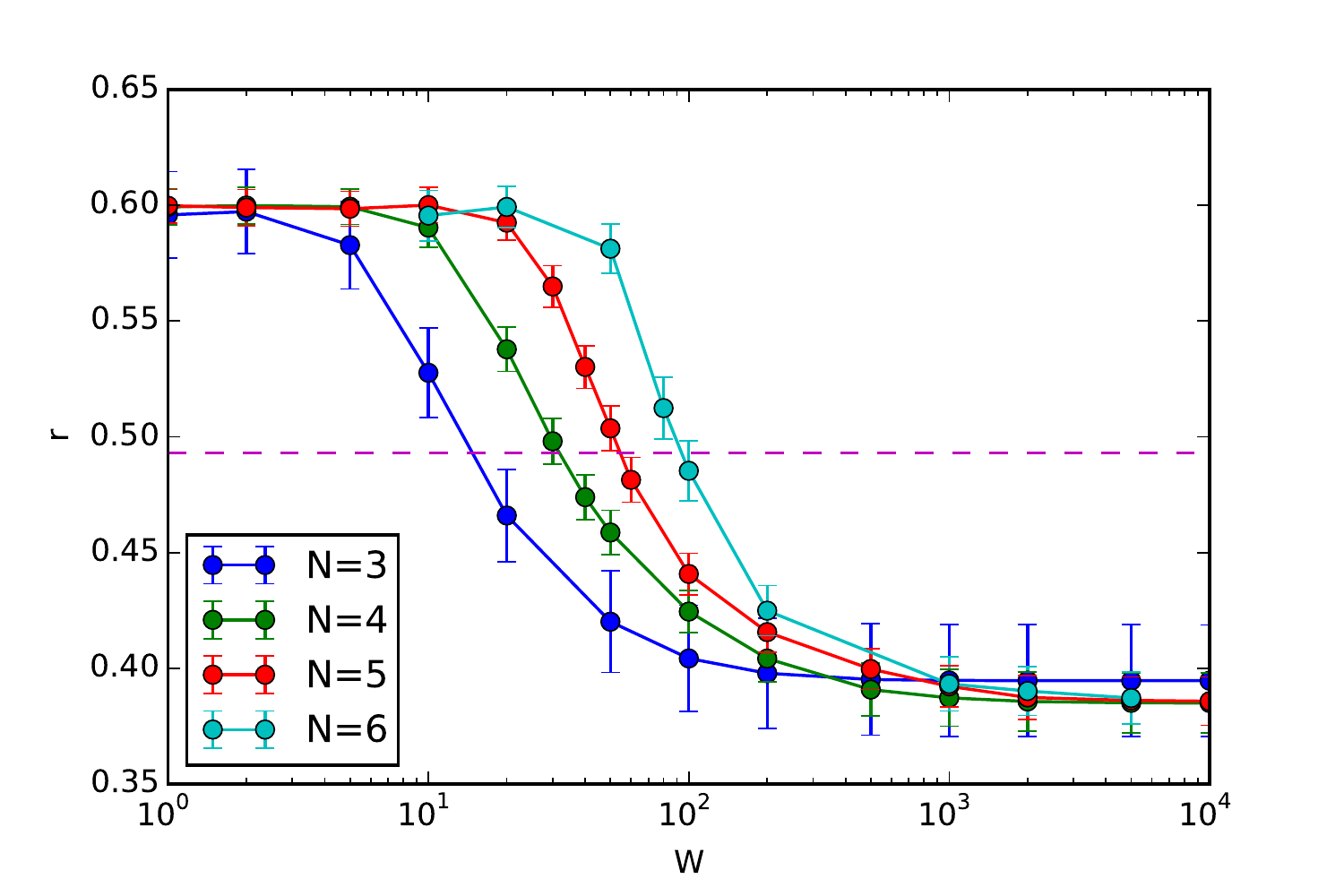}
\caption{Level spacing ratio $r$ vs. disorder strength, $W$ (Note that the x-axis is linear in the {\it logarithm} of $W$). The data clearly shows MBL-like behavior at large $W$ and extended state behavior at small $W$. However, when multiple system sizes are considered, we see that the critical $W$ for MBL increases rapidly with system size, suggesting that the system does not localize in the thermodynamic limit. Data is obtained by averaging $200$ realizations of disorder, where for each realization $100$ eigenenergies have been computed at an energy density approximately halfway between the center of the band and the lower edge. Similar results are found for other energies.
The dashed horizontal line indicates the value of $r$ halfway between the GUE and Poisson values.
\label{onethird}
}
\end{figure}

The results of measuring $r$ for a variety of number of electrons $N_e$ at filling fraction $\nu=1/3$ are shown in Fig.~\ref{onethird}. 
At small disorder we see GUE statistics, and as the disorder increases we can see a clear transition to a smaller value of $r$, indicating Poisson behavior. 
However, the location of this transition is strongly dependent upon the system size (note that the $x$-axis on this plot is logarithmic). Therefore we suspect that in the thermodynamic limit, the system displays GUE statistics everywhere.

\begin{figure}
\includegraphics[width=\linewidth]{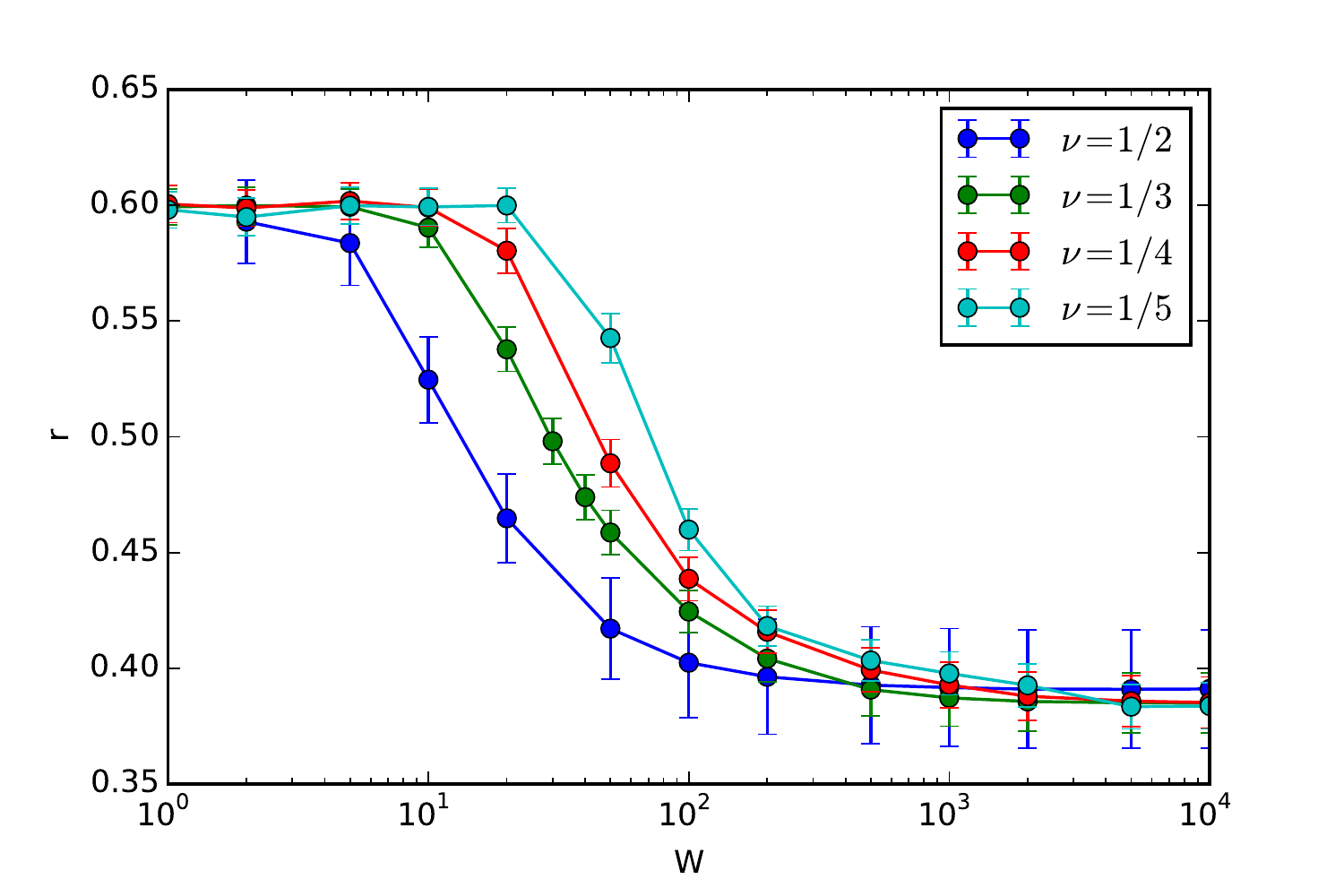}
\caption{
Same as Fig.~\ref{onethird}, but for $N_e=4$ and different filling fractions. The results are qualitatively the same regardless of filling fraction. This confirms that the transition we are probing is not related the topological properties of the ground states. The transition moves to larger disorder strength at smaller $\nu$ since the number of of flux quanta, and therefore the size of the system, is greater. }
\label{multinu}
\end{figure}

\begin{figure}
\includegraphics[width=\linewidth]{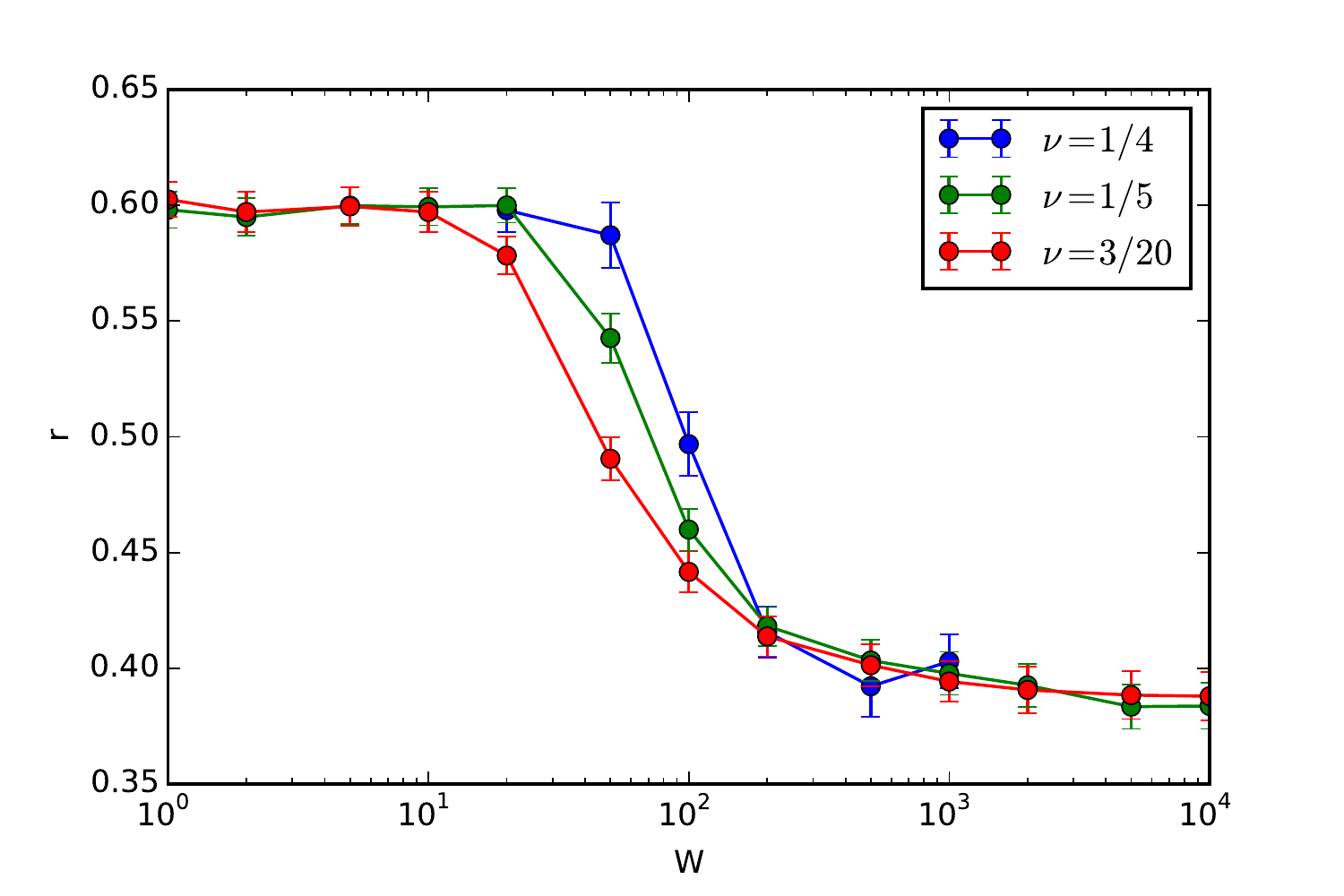}
\caption{
Same as Fig.~\ref{onethird}, but for $N_\Phi=20$ and different numbers of electrons. We see that increasing the number of electrons increases the critical disorder strength, though the effect is not as large as for increasing $N_\Phi$. }
\label{multiNPhi}
\end{figure}

Since we are interested in the properties of highly excited states, our results should not depend on the low energy properties of the system, such as whether or not the ground state exhibits the fractional quantum Hall effect. 
In Fig.~\ref{multinu} we show a similar plot as Fig.~\ref{onethird}, but for multiple filling fractions and a fixed number of electrons ($N_e=4$). 
As we suspected, the same qualitative behavior is seen regardless of filling fraction, even at fractions like $\nu=1/2$ where the ground state is very different from the one at $\nu=1/3$. 
 The location of the transition seems to display a similar dependence on system size as in Fig.~\ref{onethird}--
it moves to higher disorder at smaller fillings. This is mainly because we are working with a fixed number of electrons, and therefore smaller fillings imply more flux quanta and larger system sizes.
A better quantitative estimate of the effect of the filling fraction on the transition point can be found by comparing data with the same number of flux quanta $N_{\Phi}$, but different numbers of electrons. We show such data in Fig.~\ref{multiNPhi}. 
We observe that increasing the filling at fixed system size leads to a slightly larger critical disorder strength.

\begin{figure}
\includegraphics[width=\linewidth]{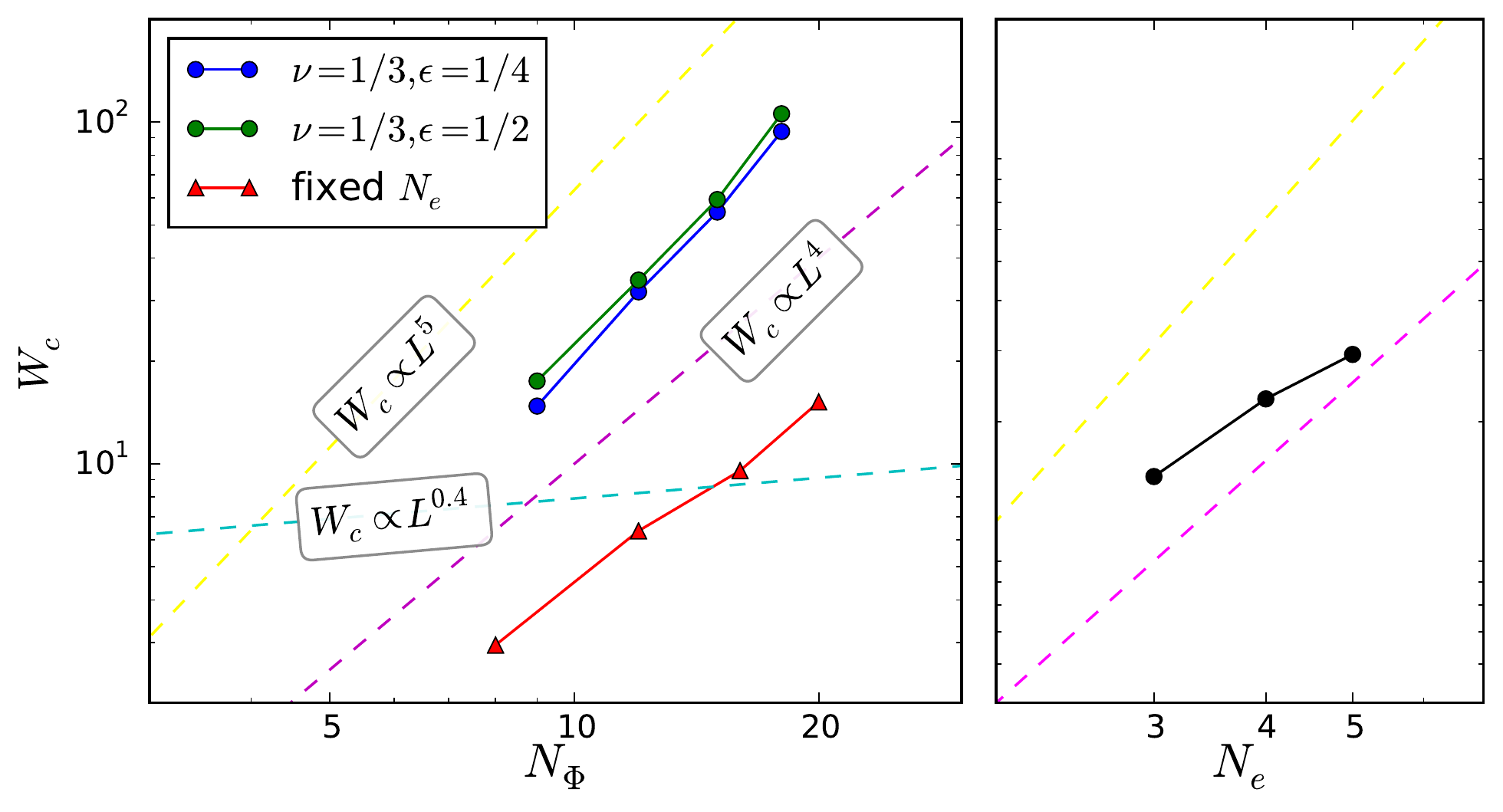}
\caption{The critical disorder strength of $W_c$, as a function of system size, plotted on a logarithmic plot. 
We define $W_c$ to be where $r$ takes a value halfway between the GUE and Poisson values (indicated by a dashed line in Fig.~\ref{onethird}).
Panel (a) shows data for different $N_{\Phi}$ holding fixed either the filling fraction or the number of electrons. For the data with fixed filling fraction $\nu=1/3$ we also show data for two different energy densities, finding very little dependence on energy density. The data with fixed $N_e$ has been offset vertically by a factor of $5$ to make the figure easier to read. The dashed lines show power laws with various exponents. It seems that the data at $\nu=1/3$ has $W\propto L^{4-5}$. The data with fixed $N_e$ is similar but grows slightly more slowly. This is consistent with the data in (b), where it can be seen that holding $N_\Phi$ fixed and increasing $N_e$ (i.e. increasing the filling  fraction) increases the critical disorder strength, but a rate slower than that for the data in (a). The dashed lines in Fig. (b) represent the same power laws as those in (a).
 }
\label{rcrit}
\end{figure}

We define the critical disorder strength, $W_c$, as the value of $W$ when $r$ is halfway between its GUE and Poisson values. Thus in Fig.~\ref{onethird}, $W_c$ is where the various curves intersect the dashed line. In Fig.~\ref{rcrit}(a) we show the dependence of $W_c$ on system size for the data in Fig.~\ref{onethird}. For the data at $\nu=1/3$ we show both data corresponding to an energy density (defined as the fraction of the distance between the lowest and highest energy eigenstates) of $\epsilon=1/4$ (corresponding to the data in Fig.~\ref{onethird}) as well as for $\epsilon=1/2$. 
 The data is plotted on a double logarithmic plot, and seems to show that the transition point increases as a large power of system size, which extrapolates to infinity in the thermodynamic limit. The exponent of this power law growth seems to be independent of energy density.
We also show in Fig.~\ref{rcrit}(a) $W_c$ for the data at fixed $N_e$, taken from Fig.~\ref{multinu}. We again find power law behavior in this case; the power seems slightly reduced compared to the data at $\nu=1/3$. Fig.~\ref{rcrit}(b) shows the $W_c$ values extracted from Fig.~\ref{multiNPhi}, where system size is held fixed and filling fraction is increased. We see that increasing filling fraction increases $W_c$ only slowly, but we expect that this increase explains the difference between the slopes in Fig.~\ref{rcrit}(a). 

The basis of Ref.~\onlinecite{PotterNandkishore}'s argument for the absence of localization is a comparison of the single-particle level spacing at some distance $R$, with the size of the tunneling elements (which are enhanced by the  existence of the topologically protected delocalized state) between two states separated by $R$. The latter is claimed to be as large as:
\begin{equation}
T\sim V^2 R^{-\frac12 (d+1/\nu)}
\label{tunneling}
\end{equation}
where $V$ is the strength of the interaction, and $d\equiv2$ is the dimensionality of the system. Since the level spacing is proportional to $R^{-d}$, the condition for localization is $\nu d < 1$, since if that is not the case then there is tunelling between states at large $R$.

Are our observations consistent with these arguments? Let us define $L$ to be the linear dimension of the system (in units of magnetic length). In this work $L$ is approximately $8-11$ magnetic lengths. Since $L$ is the longest length scale in the problem we replace $R$ with $L$ in Eq.~(\ref{tunneling}). This tells us that the transition from localized to delocalized happens when
\begin{equation}
V \sim L^{\frac {1}{4\nu} (1-\nu d)}.
\label{VvsL}
\end{equation}
In our projected problem the only energy scales are interaction strength and disorder, so we can replace $V$ with $W^{-1}$. Since $\nu\approx 2.5$, this leads to the prediction that the critical disorder strength should be $W_c\propto L^{\approx 0.4}$. 
In Fig.~\ref{rcrit}, we do seem to observe a power law dependence of the transition point, but we find $W_c\propto L^{4-5}$, an exponent which is an order of magnitude larger. 

\section{Discussion}
We have studied the behavior of the finite energy density eigenstates of interacting electrons in a single Landau level using the shift-invert method to numerically extract the many-body eigenstates for finite size systems with disorder using several hundred members of the ensemble. Our results show that although only a vanishingly small fraction of the single particle states are effectively delocalized in the thermodynamic limit for non-interacting electrons, the addition of electron-electron interactions causes delocalization to spread though the entire many-body spectrum. 

Our main result thus agrees with the analytical calculations of Ref.~\onlinecite{PotterNandkishore}, and therefore suggests that the basic assumptions made in that work are justified.
However, the scaling of the critical disorder with size is found to be much more dramatic. This could be because of the small sizes that we were able to study. However, it could also be that mechanisms other than those considered in Ref.~\onlinecite{PotterNandkishore} may dominate. In this context, it is worth mentioning that related studies of MBL and topological transitions have found similar behavior, including violation of the Harris-CCFS bound.\cite{Alet2015,Liu2016} 

Our work has implications for the localization properties of other systems which have single particle mobility edges. Other groups\cite{Modak2016,Li2016} have claimed, in the context of a quasi-periodic one dimensional model with localized and extended single particle states (both of which scale with system size), that with the introduction of interactions, MBL can still occur. Our work is for a two-dimensional system with quenched disorder with a  {\it sub-thermodynamic} (but divergent in the thermodynamic limit) number of single-particle eigenstates. Nevertheless, in the thermodynamic limit, this subthermodynamic number of states appear sufficient to destroy many-body localization. The stark contrast between the two results begs further clarification.

Our work is also one of the few numerical studies of many-body localization in two dimensions. Recent work\cite{Huveneers2016} has proposed that MBL may be inherently unstable in 2D, and this might also serve as an explanation of our results. 
The instability discussed in Ref.~\onlinecite{Huveneers2016} is due to large rare regions. The probability of having such a rare region in our finite system is propotional to the system size, i.e. $\propto L^2$. This is also power law growth, though with a much smaller exponent than we observe. Note however that a previous study of a two-dimensional transverse-field Heisenberg model\cite{chain} observed a growth of critical disorder strength with an exponent $\approx 2$. This suggests that the enhanced exponent observed in this work is related to the topological nature of the Landau level.


{\bf Acknowledgements:}
We acknowledge helpful conversations with Rahul Nandkishore,  and Zhao Liu. This work was supported by Department of Energy  BES Grant DE-SC0002140. The matrix decomposition step of the shift-invert method was accomplished using the SuperLU library.\cite{SuperLU1,*SuperLU2,*SuperLU3}

\bibliography{QH_MBL}
\end{document}